\documentclass[twocolumn,prl,amsmath,amssymb,superscriptaddress]{revtex4-2}

\usepackage[pdftex]{graphicx}
\usepackage{dcolumn}
\usepackage{bm}
\usepackage{amsmath, amssymb,amsfonts}
\usepackage{mathrsfs}
\usepackage{physics}
\usepackage{type1cm}
\usepackage[colorlinks,linkcolor=blue,citecolor=blue]{hyperref}

\begin{document}

\title{
Melting of a Bosonic Mott Insulator in Kagome Optical Lattices with Sign-Inverted Hopping
}

\author{Kazuma Nagao}
\email{kazuma.nagao@riken.jp}
\affiliation{%
Computational Materials Science Research Team, RIKEN Center for Computational Science (R-CCS), Hyogo 650-0047, Japan
}%
\affiliation{%
Quantum Computational Science Research Team, RIKEN Center for Quantum Computing (RQC), Wako, Saitama 351-0198, Japan
}%
\author{Daisuke Yamamoto}
\affiliation{%
Department of Physics, College of Humanities and Sciences, Nihon University, Sakurajosui, Setagaya, Tokyo 156-8550, Japan
}%
\affiliation{%
Quantum Many-Body Dynamics Research Team, RIKEN Center for Quantum Computing (RQC), Wako, Saitama 351-0198, Japan
}%
\author{Seiji Yunoki}
\affiliation{%
Computational Materials Science Research Team, RIKEN Center for Computational Science (R-CCS), Hyogo 650-0047, Japan
}%
\affiliation{%
Quantum Computational Science Research Team, RIKEN Center for Quantum Computing (RQC), Wako, Saitama 351-0198, Japan
}%
\affiliation{%
Computational Condensed Matter Physics Laboratory, RIKEN Pioneering Research Institute (PRI), Saitama 351-0198, Japan
}%
\affiliation{%
Computational Quantum Matter Research Team, RIKEN Center for Emergent Matter Science (CEMS), Saitama 351-0198, Japan
}%
\author{Ippei Danshita}
\affiliation{%
Department of Physics, Kindai University, 3-4-1 Kowakae, Higashi-Osaka, Osaka 577-8502, Japan
}%

\date{\today}

\begin{abstract}

Using the discrete truncated Wigner approximation (dTWA), we investigate the nonequilibrium dynamics of ultracold bosons confined in optical kagome lattices, focusing on both unfrustrated positive and frustrated negative hopping regimes. 
We consider a protocol in which the system is initialized in a Mott insulating state at unit filling, and the hopping amplitude is gradually increased from zero.  
For positive hopping, the melting of the Mott insulator is accompanied by the emergence of a sharp peak in the momentum distribution at the $\Gamma$ point of the lowest band, signaling the onset of superfluidity.   
In contrast, for negative hopping, the Mott insulator melts into a highly nontrivial state without long-range phase coherence, characterized instead by a broad momentum distribution within the flat band, consistent with recent experimental observations. 
These results demonstrate the applicability of dTWA to highly frustrated quantum systems and offer a new route for numerically exploring the dynamics of frustrated quantum magnets.

\end{abstract}

\maketitle

Geometrical frustration in antiferromagnets on nonbipartite lattices, such as triangular and kagome lattices, has been at the center of the search for intriguing many-body phenomena in condensed-matter physics~\cite{moessner2006geometrical, lewenstein2007ultracold, batista2016frustration}.
A wide range of candidate materials exhibiting such nonbipartite geometries have been extensively investigated, including CsCuCl$_3$~\cite{nojiri1988magnetic, yamamoto2021continuous, nihongi2024field}, Ba$_3$CoSb$_2$O$_9$~\cite{shirata2012experimental, zhou2012successive, susuki2013magnetization, yamamoto2014quantum, yamamoto2015microscopic}, ZnCu$_3$(OH)$_6$Cl$_2$~\cite{mendels2010quantum, khuntia2020gapless}, and YCu$_3$(OD)$_{6+x}$Br$_{3-x}$~\cite{chen2020quantum, jeon2024one, zheng2025unconventional}. 
Because simulations of frustrated quantum magnets are notoriously difficult on classical computers, such systems are regarded as promising targets for quantum simulation using ultracold atoms~\cite{lewenstein2007ultracold}. 
Indeed, a number of cold-atom experiments have successfully realized frustrated models on nonbipartite lattices, including the Ising~\cite{semeghini2021probing}, XY~\cite{struck2011quantum, struck2013engineering}, Heisenberg~\cite{xu2023frustration}, and Bose-Hubbard models~\cite{ozawa2023observation, donini2025bose}.

In the Bose-Hubbard model, as realized with ultracold Bose gases in optical lattices~\cite{fisher1989boson, greiner2002quantum}, a negative hopping integral plays a role analogous to antiferromagnetic coupling in spin systems and is a key source of frustration. 
Recent experiments~\cite{donini2025bose} have employed ultracold gases at negative absolute temperatures~\cite{braun2013negative, yamamoto2020frustrated} to realize sign-inverted hopping amplitudes, thereby implementing frustrated Bose-Hubbard systems at low filling, where quantum fluctuations are expected to be significant. 
In particular, they have investigated the dynamical melting of an initial Mott insulator (MI) at unit filling under a slow ramp-down of the lattice depth in a kagome lattice. 
The kagome geometry is especially interesting due to the strong interplay of quantum and thermal fluctuations with geometrical frustration, which has led to theoretical predictions of various exotic phases, such as flat band Bose-Einstein condensates~\cite{you2012superfluidity,julku2021quantum}, supersolids~\cite{huber2010bose}, and trion superfluids~\cite{you2012superfluidity}. 
In the experiment of Ref.~\cite{donini2025bose}, it was found that after the ramp, the MI melts into a state where most atoms are broadly distributed within the flat band in momentum space, indicating the absence of superfluid (SF) order. 
This behavior stands in sharp contrast to the case of positive hopping, where a sharp peak emerges at the $\Gamma$ point of the lowest band in the momentum distribution~\cite{jo2012ultracold}. 
This experimental observation presents a theoretical challenge, as the nonequilibrium dynamics of frustrated quantum systems on kagome lattices remain notoriously difficult to calculate and understand.

\begin{figure*}
\begin{center}
\includegraphics[width=\textwidth,keepaspectratio]{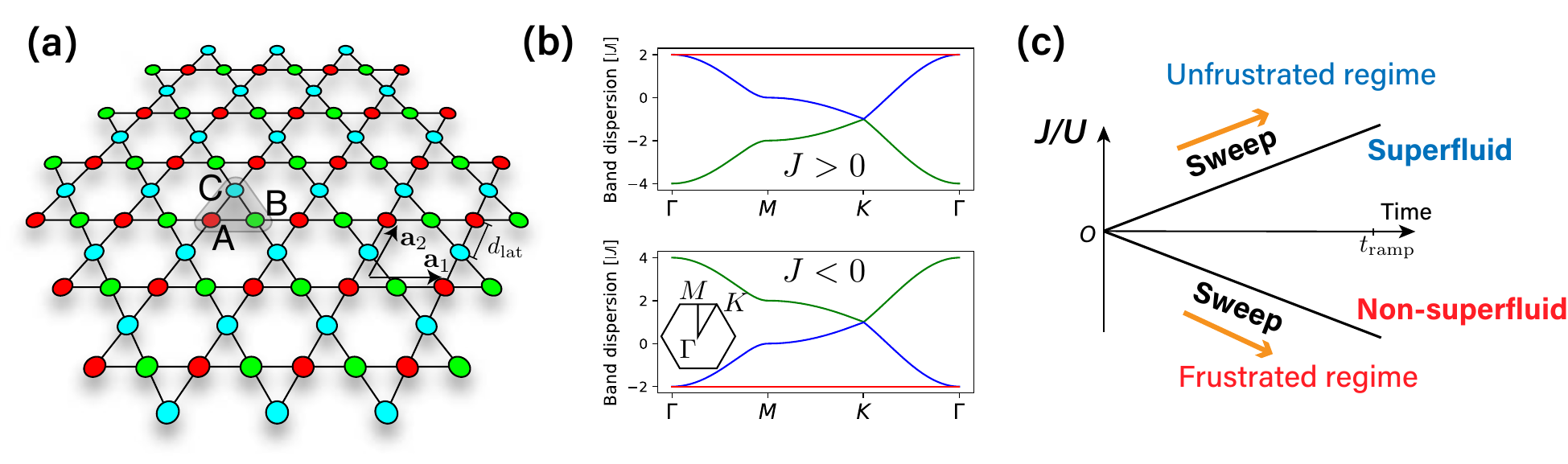}
\vspace{-6mm}
\caption{
(a)
Kagome optical lattice.
The unit cell, consisting of three sublattices ($A$, $B$, and $C$), is spanned by the lattice vectors 
${\bf a}_1 = 2d_{\rm lat}(1,0)^{\rm T}$ 
and 
${\bf a}_2 = d_{\rm lat}(1,\sqrt{3})^{\rm T}$, where $d_{\rm lat}$ is the lattice constant. 
The sublattice positions are given by ${\bf r}_{A} = {\bf R}_{mn}$, ${\bf r}_{B} = {\bf R}_{mn}+{\bf a}_{1}/2$, and ${\bf r}_{C} = {\bf R}_{mn}+{\bf a}_{2}/2$, where ${\bf R}_{mn}  \in \{ m {\bf a}_1 + n {\bf a}_2 | m,n \in {\mathbb Z} \}$.
(b)
Single-particle band structure for the hopping term in Eq.~(\ref{eq:hamiltonian}).
The upper panel shows the case of positive hopping ($J>0$), where the flat band appears at the top of the spectrum.
In the lower panel, the sign inversion of $J$ flips the band structure, placing the flat band at the bottom. 
The inset shows the first Brillouin zone (BZ) of the kagome lattice. 
(c)
Sweep protocol of $J(t)/U$. 
The slope of the linear ramp determines the sweep rate. 
The ramp ends at time $t=t_{\rm ramp}$.
}
\label{fig:1}
\end{center}
\end{figure*}

In this paper, we study the dynamics of the Bose-Hubbard model on a kagome lattice starting from a MI at unit filling, under a slow ramp of the hopping amplitude, using the SU($n$) discrete truncated Wigner approximation (dTWA)~\cite{schachenmayer2015many,davidson20153,wurtz2018cluster,zhu2019generalized,nagao20213,kunimi2021performance,schultzen2022semiclassical,nagao2024discrete,braemer2024cluster,nagao2025two}, which incorporates quantum fluctuations in the strongly interacting regime. 
We show that while the MI melts into a SF state in the case of positive hopping, it evolves into a state without long-range SF order in the case of negative hopping, characterized by a broad momentum distribution over the flat band and rapidly decaying single-particle correlations in real space. 
This behavior is in qualitative agreement with recent experimental observations~\cite{donini2025bose}. 
We also analyze the triangular-lattice case to further examine the qualitative validity of the SU($n$) dTWA for frustrated Bose-Hubbard systems.

{\it Model.---}
We consider the Bose-Hubbard Hamiltonian describing bosons confined in an optical kagome lattice~\cite{jo2012ultracold}:
\begin{align}
{\hat {\cal H}} = - \sum_{i,j} J_{ij} {\hat b}^{\dagger}_{i}{\hat b}_{j} + \frac{U}{2}\sum_{i} {\hat b}^{\dagger}_{i}{\hat b}^{\dagger}_{i} {\hat b}_{i}{\hat b}_{i}, \label{eq:hamiltonian}
\end{align}
where $J_{ij}$ denotes the hopping amplitude, which takes a nonzero value $J$ only for nearest-neighbor pairs, and ${\hat b}^\dag_i$ is the boson creation operator at site $i$. 
The kagome lattice is a tripartite structure consisting of $A$, $B$, and $C$ sublattices [Fig.~\ref{fig:1}(a)].
A key feature of this system is the presence of a flat band at the bottom of the non-interacting single-particle spectrum for negative hopping amplitudes $J < 0$~\cite{huber2010bose,you2012superfluidity}, as shown in Fig.~\ref{fig:1}(b).
The band dispersions in this case are given by 
\begin{align}
\hbar \omega_{0} = -2|J|, \;\; \hbar \omega_{\pm}({\bf k}) = |J| [ 1 \pm \sqrt{3 + 2 \Lambda({\bf k})}],
\end{align}
where $\Lambda({\bf k}) = \cos(k_1) + \cos(k_2) + \cos(k_2 - k_1)$ and $k_{s=1,2} = {\bf k} \cdot {\bf a}_{s}$ with ${\bf a}_{s}$ being the lattice vectors.
The flatness of $\hbar \omega_{0}$ originates from the combination of local frustration of bosonic phase coherence on each triangular plaquette--i.e., the shaded region in Fig.~\ref{fig:1}(a)--and destructive interference across connected plaquettes, which inhibits boson delocalization. 
In the non-interacting limit, the ground state is macroscopically degenerate in momentum space. 
When the sign of $J$ is inverted from negative to positive, the lowest band becomes $\hbar \omega_{+}({\bf k})$, which has a minimum at the $\Gamma$ point, while the flat band is pushed to the top of the spectrum, as also shown in Fig.~\ref{fig:1}(b). 
Therefore, at zero temperature, free bosons condense at the $\Gamma$ point of the first Brillouin zone (BZ).

Our interest lies in the properties of low-temperature states of Eq.~(\ref{eq:hamiltonian}) for finite interaction $U>0$, which can be probed by a slow sweep of the optical-lattice depth starting from a MI ground state~\cite{braun2013negative, donini2025bose}. 
Previous theoretical studies have explored the possible low-temperature phases of Eq.~(\ref{eq:hamiltonian}) using various analytical and numerical methods~\cite{huber2010bose,you2012superfluidity,zhang2015tunable,julku2021quantum}. 
For positive hopping, quantum Monte Carlo (QMC) analyses~\cite{zhang2015tunable} have shown that the system exhibits three distinct phases at low temperatures, i.e., SF, MI, and normal-fluid phases, as is also the case in Bose-Hubbard models on bipartite lattices~\cite{fisher1989boson}. 
In contrast, for negative hopping, where QMC is hindered by the sign problem, Bogoliubov analyses--valid in the weakly-interacting, high-density regime--predict that the ground state is a Bose-condensed state at the $K$-point of the flat band, while the low-temperature thermal state is a trion superfluid~\cite{you2012superfluidity}. 
However, theoretical analyses in the strongly interacting regime relevant to recent experiments~\cite{donini2025bose} remains limited. We therefore employ the SU($n$) dTWA to study the sweep dynamics in this strongly interacting regime.

\begin{figure*}
\begin{center}
\includegraphics[width=\textwidth,keepaspectratio]{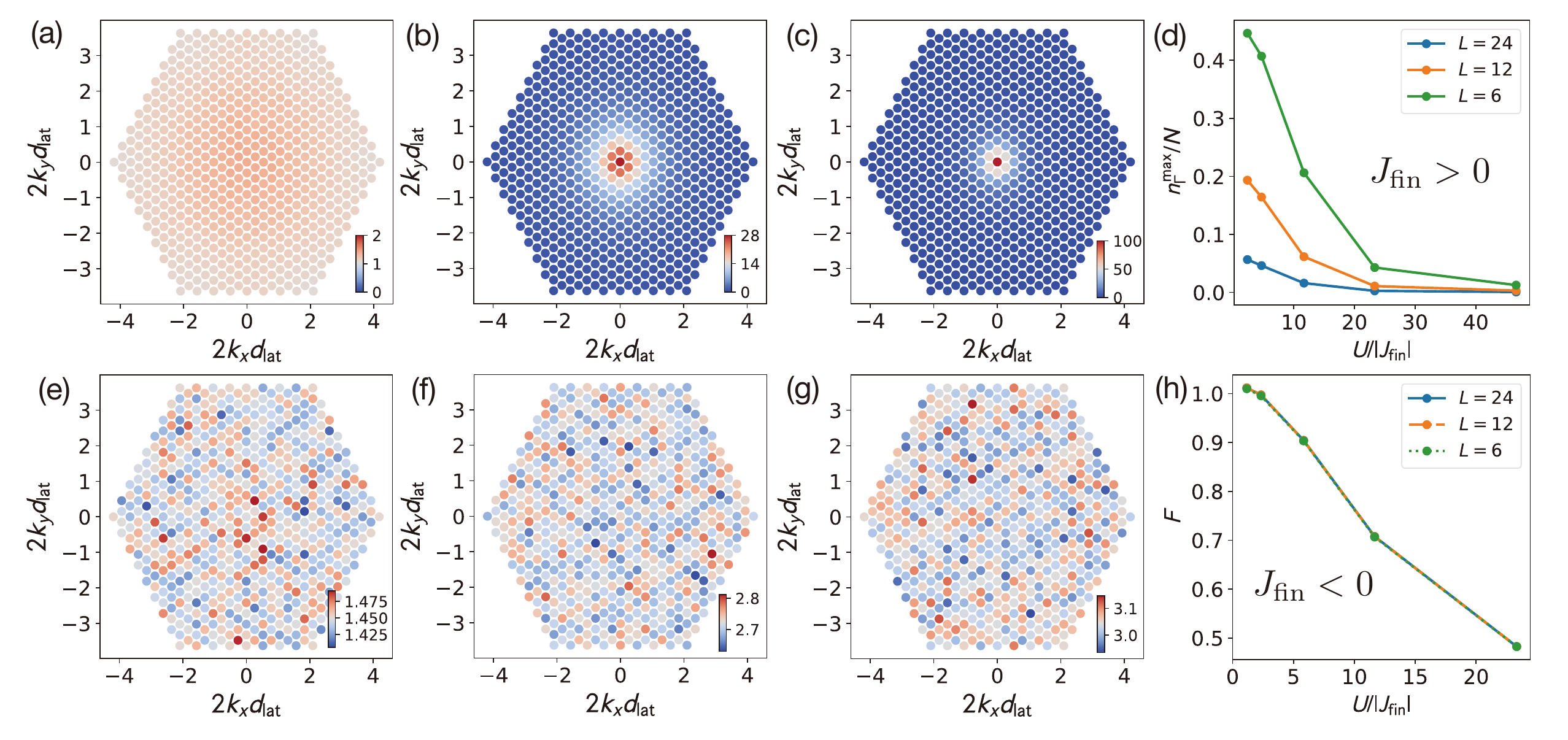}
\vspace{-7mm}
\caption{
Effective lowest-band distribution $n_{\bf k}^{\rm max}$ in the final states of the kagome Bose-Hubbard model after the sweep dynamics. 
(a-c)
$\Gamma$-point condensation in dynamically prepared states with unfrustrated hopping amplitudes: (a) $g_{\rm fin} = 0.5$, (b) $2$, and (c) $10$, for $N_{\rm s} = 1728$ sites and $L=24$. 
(d) Condensate fraction at the $\Gamma$ point as a function of $U/|J_{\rm fin}|$ for $J_{\rm fin} > 0$ and system sizes $L=6$, $12$, and $24$. 
(e-g)
Broad, flat distribution over the BZ in the lowest band for frustrated hopping amplitudes: (e) $g_{\rm fin} = -0.5$, (f) $-2$, and (g) $-10$, with $N_{\rm s} = 1728$ sites and $L=24$.
The fluctuations in the distribution arise from statistical sampling of trajectories in the dTWA.
(h)
Fraction $F = N^{-1} \sum_{{\bf k} \in {\rm B.Z.}} n^{\rm max}_{\bf k}$, defined as the average of $n^{\rm max}_{\bf k}$ over the first BZ for $J_{\rm fin} < 0$, increases with $|J_{\rm fin}|/U$. 
}
\label{fig:2}
\end{center}
\end{figure*}

To this end, we simulate a linear ramp of the hopping amplitude starting from a unit-filling MI state, which is the ground state at $J/U = 0$. 
As shown in Fig.~\ref{fig:1}(c), the time dependence is given by $J(t) = g_{\rm fin}|J_{\rm c}|(t/t_{\rm ramp})$, where $t_{\rm ramp}$ is the total duration of the ramp, $g_{\rm fin} = J_{\rm fin}/|J_{\rm c}|$, and $J_{\rm fin} = J(t = t_{\rm ramp})$.
The sign of $g_{\rm fin}$ determines the regime: positive for the unfrustrated case ($J>0$) and negative for the frustrated case ($J<0$). 
We assume that the onsite interaction strength $U$ remains constant during the evolution.
The critical ratio $U/|J_{\rm c}|$ for the SF-MI transition at zero temperature has been estimated using single-site Gutzwiller mean-field theory~\cite{krauth1992gutzwiller,sheshadri1993superfluid,yamamoto2020frustrated}. 
For the kagome lattice with coordination number $z=4$, the mean-field critical values are $U/|J_{\rm c}| \approx 23.3$ for $J >0$ and $11.7$ for $J <0$. 
Note, however, that in the $J<0$ case, the Gutzwiller theory fails to capture even qualitative features of the system due to strong quantum fluctuations and frustration. 
We thus regard $U/|J_{\rm c}|$ as a reference value indicating the approximate location of the SF-MI transition in the absence of quantum fluctuation effects. In the following calculations, we set $t_{\rm ramp} z|J_{\rm c}| / \hbar = 2000$ to ensure quasi-adiabatic dynamics~\cite{sm} and assume periodic boundary conditions along the $\bm a_1$ and $\bm a_2$ directions for a system with $N_{\rm s}=3L^2$ sites, where $L$ is the linear dimension of the system.

{\it Fraction in the effective lowest band.---}
To characterize the growth of phase coherence after the sweep, we analyze the one-body reduced density matrix (OBRDM), defined as $C_{{\bf R}\alpha,{\bf R}' \beta}(t) = \langle {\hat b}^{\dagger}_{{\bf R}\alpha} (t) {\hat b}_{{\bf R}' \beta} (t) \rangle$, 
where ${\hat b}_{{\bf R}\alpha}$ is the bosonic annihilation operator at sublattice site $\alpha$ in the unit cell located at ${\bf R}$.
We numerically solve the time-dependent Schr\"odinger equation for the kagome Bose-Hubbard model using the SU($n$) dTWA~\cite{schachenmayer2015many,davidson20153,wurtz2018cluster,zhu2019generalized,nagao20213,kunimi2021performance,nagao2024discrete,nagao2025two}. 
In this method, fluctuating initial-state configurations are sampled at $t=0$ in phase space and propagated according to a nonlinear equation of motion, which approximates the quantum dynamics by an ensemble of mean-field trajectories~\cite{blakie2008dynamics,polkovnikov2010phase}. 
To minimally incorporate soft-core onsite interactions, we reduce Eq.~(\ref{eq:hamiltonian}) to a three-state truncated pseudospin model~\cite{huber2007dynamical,nagao2018response}, where the maximum site occupancy is restricted to two. 
The bosonic field operators are then represented by finite-dimensional matrices forming the SU(3) Lie algebra~\cite{davidson20153,nagao20213}, 
which enables the application of the SU(3) dTWA for the subsequent analysis.
Further details are provided in the Supplemental Material (SM)~\cite{sm}.
Note that since higher-occupancy states such as triplons are excluded, possible trion superfluid~\cite{you2012superfluidity} is not captured in the present study.

The relevant quantity for identifying condensation in the lowest band is the distribution of the maximum eigenvalues of the OBRDM. 
By performing a Fourier transformation over ${\bf R}$ and ${\bf R}'$, the OBRDM becomes block diagonal form in momentum space. 
The intra-cell correlations at a given momentum ${\bf k}$ are represented as a $3 \times 3$ matrix: 
\begin{align}
{\cal C}^{\alpha \beta}_{\bf k} = \frac{1}{L^2}\sum_{{\bf R},{\bf R}'} C_{{\bf R}\alpha,{\bf R}' \beta} e^{i({\bf R} - {\bf R}')\cdot {\bf k}} e^{i({\bf u}_{\alpha} - {\bf u}_{\beta})\cdot {\bf k}}. 
\end{align}
A unitary transformation ${\cal U}_{\bf k}$ diagonalizes ${\cal C}_{\bf k}$ as ${\cal U}^{\dagger}_{\bf k} {\cal C}_{\bf k} {\cal U}_{\bf k} \rightarrow \sum_{s=0,1,2} {\bm \phi}_{s} ({\bf k})  n^{s}_{\bf k} {\bm \phi}_{s}^{\dagger} ({\bf k})$, 
with eigenvalues ordered as $n^{0}_{\bf k} > n^{1}_{\bf k} > n^{2}_{\bf k}\, (\ge 0)$. 
The eigenvectors ${\bm \phi}_{s}(\bf k) $ define effective one-body states labeled by $({\bf k}, s)$, incorporating many-body offsite correlations. 
Condensation in the SF regime is identified by the momentum distribution $n^{\rm max}_{\bf k} \equiv n^{0}_{\bf k}$.

We first consider a slow sweep into the unfrustrated regime of Eq.~(\ref{eq:hamiltonian}), where the final state is expected to approach a correlated ground state with positive hopping, $J_{\rm fin} > 0$.
As shown in Figs.~\ref{fig:2}(a)-\ref{fig:2}(c), the momentum distribution of the maximum eigenvalues exhibits the formation of a peak at the $\Gamma$ point as $g_{\rm fin}$ increases.  
This indicates that the ground state in the unfrustrated regime is a $\Gamma$-point SF. 
The condensate fraction is defined as $n^{\rm max}_{\Gamma}/N$, where $N$ is equal to the total number of particles in the initial state at unit filling.
The system-size dependence of the condensate fraction is shown in Fig.~\ref{fig:2}(d).
While the condensate fraction decreases with increasing system size, the prominent peak structure near zero momentum persists even for large $L$, indicating the development of long-range SF correlations. 
We note that peak formation at the $\Gamma$ point remains visible even when triple occupation is included, as verified using the SU($4$) dTWA formulation~\cite{sm}.

Interestingly, 
inverting the sign of the hopping amplitude to enter the frustrated regime leads to a drastic change in the behavior of the final state.
As shown in Figs.~\ref{fig:2}(e)-\ref{fig:2}(g), the occupation in the lowest single-particle band becomes broadly distributed over the entire first BZ, with no peak formation observed at any interaction strength. 
Therefore, the initially prepared MI melts into a disordered state during the sweep. 
To quantify the occupation of the lowest band, we define a global fraction over the flat band as $F = N^{-1} \sum_{{\bf k} \in {\rm B.Z.}} n^{\rm max}_{\bf k}$.
As shown in Fig.~\ref{fig:2}(h), $F$ increases with $|J_{\rm fin}|/U$ and shows negligible dependence on system size. 
This result suggests that in the weakly interacting regime, the flat band becomes dominantly occupied while long-range phase coherence is suppressed
\footnote{
Note that although the total particle number is conserved 
as $N = \sum_{{\bf R},\alpha}C_{{\bf R}\alpha,{\bf R}\alpha}$ 
in the simulations, the fraction $F$ can exceed unity for $J_{\rm fin}/J_{\rm c}\gtrsim 10$.
This unphysical artifact in the strongly nonlinear regime is attributed to negative-valued populations in the higher bands within the accuracy limits of dTWA. However, these contributions remain subdominant compared to the lowest-band occupation. 
}.

{\it Real-space correlations.---}
Next, we analyze site-to-site correlations within each sublattice in the dynamically prepared states.
Without loss of generality, we focus on $AA$ correlations. 
We define the spatially averaged correlation function within the $A$ sublattice as ${\overline {C_{AA}}}(r) \equiv {\cal N}(r)^{-1}\sum_{|| {\bf R} - {\bf R}' ||_2 = r} C_{{\bf R}A, {\bf R}' A}$, where the summation is taken over all pairs of $({\bf R},{\bf R}')$ separated by a Euclidean distance $r$.
The factor ${\cal N}(r)$ denotes the number of such pairs.

\begin{figure}
\begin{center}
\includegraphics[width=\columnwidth,keepaspectratio]{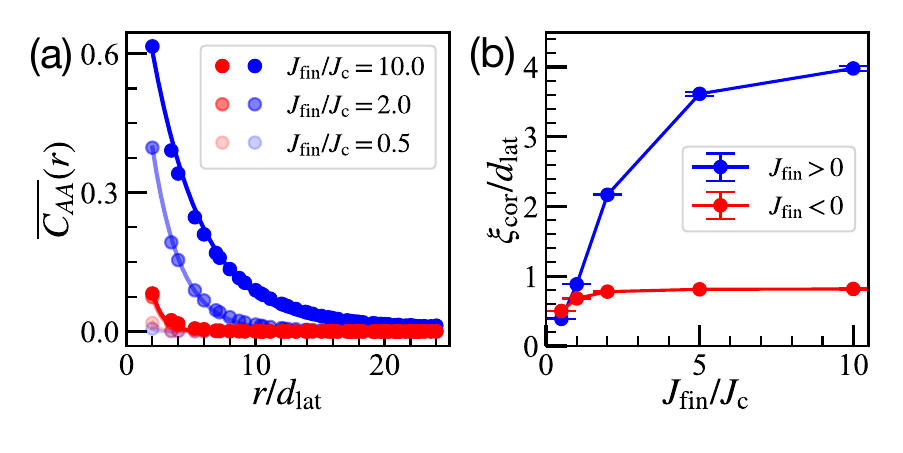}
\vspace{-8mm}
\caption{
Real-space correlation functions ${\overline {C_{AA}}}(r)$ on the kagome lattice. 
(a)
Comparison between the frustrated and unfrustrated cases for $L=24$.
Blue (red) points correspond to the results shown in Figs.~\ref{fig:2}(a)-\ref{fig:2}(c) [Figs.~\ref{fig:2}(e)-\ref{fig:2}(g)] for $J_{\rm fin} > 0$ ($J_{\rm fin} < 0$). 
Solid lines represent exponential fits of the form $e^{-r / \xi_{\rm cor}}$, where $\xi_{\rm cor}$ is the fitting parameter. 
(b) 
Correlation length $\xi_{\rm cor}$, in units of $d_{\rm lat}$, plotted as a function of $J_{\rm fin}/J_{\rm c}$.
Blue (red) points with error bars indicate the extracted values of $\xi_{\rm cor}$ for $J_{\rm fin} > 0$ ($J_{\rm fin} < 0$).
}
\label{fig:3}
\end{center}
\end{figure}

The results for ${\overline {C_{AA}}}(r)$ are shown in Fig.~\ref{fig:3}(a).
In both regimes, the correlation functions exhibit exponential decay of the form ${\overline {C_{AA}}}(r) \approx \exp(-r / \xi_{\rm cor})$. 
The extracted correlation lengths $\xi_{\rm cor}$ for both cases are plotted in Fig.~\ref{fig:3}(b).
For positive hopping, the correlation length increases with $J_{\rm fin}/J_{\rm c}$ and reaches up to $\xi_{\rm cor} \approx 4 d_{\rm lat}$ for typical values of $J_{\rm fin}$. 
Although a finite condensate fraction is observed at the $\Gamma$ point in Figs.~\ref{fig:2}(b) and \ref{fig:2}(c), the real-space correlations remain short ranged, lacking even quasi-long-range order. This suppression of coherence may be attributed to strong quantum fluctuations enhanced by the geometry of the kagome lattice, which is 
characterized by a small coordination number ($z=4$) and vertex-sharing triangular plaquettes.
In contrast, for negative hopping, the correlation length never exceeds the lattice constant, i.e., $\xi_{\rm cor} \lesssim d_{\rm lat}$. 
This result is consistent with recent experimental observations~\cite{donini2025bose}, where similar exponential decay of atomic correlations was reported in the frustrated regime.

\begin{figure}
\begin{center}
\includegraphics[width=\columnwidth,keepaspectratio]{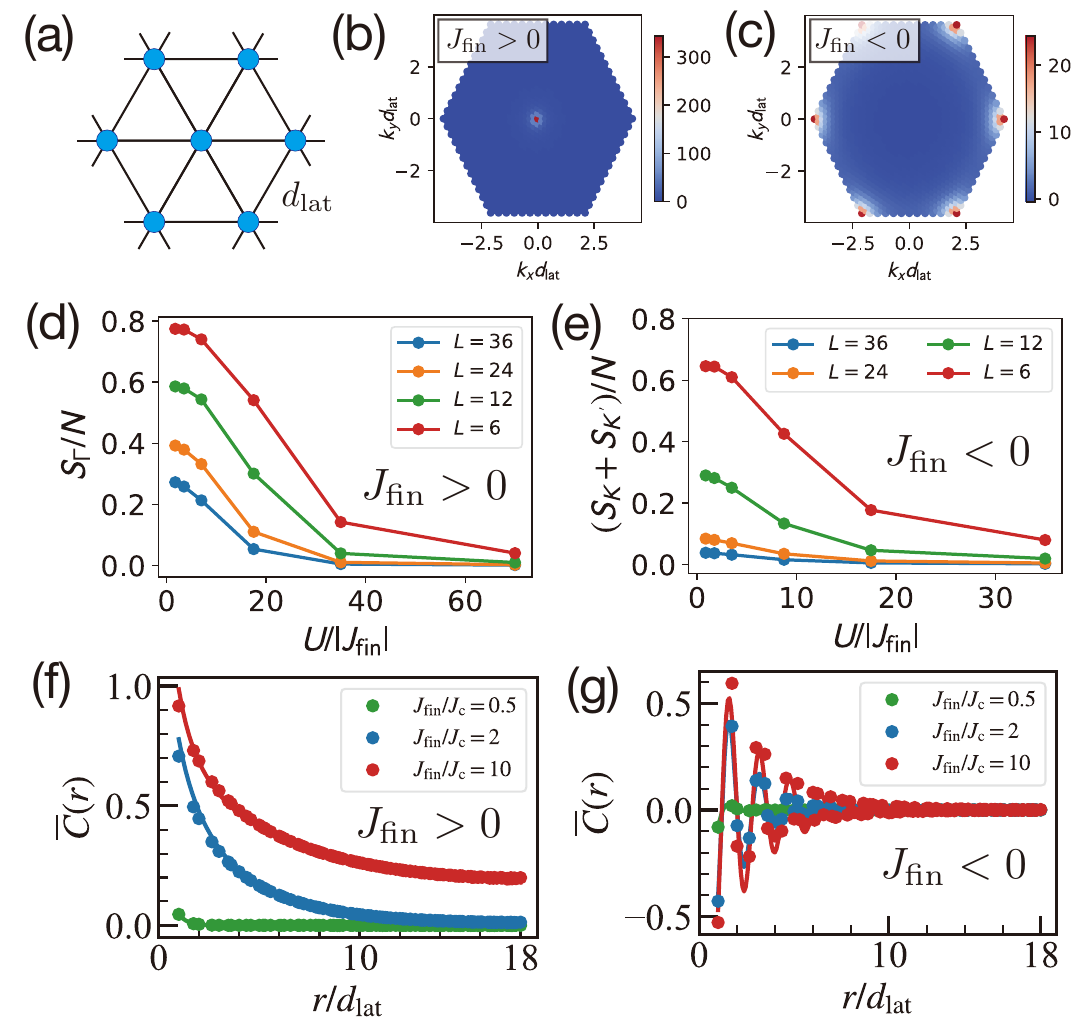}
\vspace{-5mm}
\caption{
Condensation in the triangular Bose-Hubbard model after the sweep dynamics.
(a)
Geometry of the triangular lattice. 
(b)
$\Gamma$-point condensation for positive hopping ($J_{\rm fin} > 0$). 
(c)
Condensation at the $K$ and $K'$ point for negative hopping ($J_{\rm fin} < 0$). 
(d)
Condensate fraction at the $\Gamma$ point for $J_{\rm fin} > 0$, plotted as a function of $U/|J_{\rm fin}|$.
(e)
Same as (d), but for the condensate fraction at the $K$ and $K'$ points for $J_{\rm fin} < 0$.
The vertical axis shows the sum of the contributions from ${\bf k} = {\bf Q}_{\rm K}$ and ${\bf k} = {\bf Q}_{\rm K'}$.
(f)
Real-space correlation functions ${\overline C}(r)$ for $J_{\rm fin}>0$. 
For $J_{\rm fin}/J_{\rm c} = 10$, the correlations decay as a power law with exponent $\eta \approx 0.5$.
Solid lines represent fits to the numerical data (solid circles). 
(g)
Same as (f), but for $J_{\rm fin} < 0$.
The correlations decay exponentially with oscillations. 
}
\label{fig:4}
\end{center}
\end{figure}

{\it Comparison with the triangular lattice.---}
To validate the reliability of our results, we apply the same method to a triangular-lattice Bose-Hubbard system [Fig.~\ref{fig:4}(a)] for comparison. 
Figures~\ref{fig:4}(b) and \ref{fig:4}(c) show the momentum distribution of the OBRDM, $S_{\bf k} = {\rm F.T.}[\langle {\hat b}^{\dagger}_{i}{\hat b}_{j} \rangle]$, in the triangular lattice at the end of the hopping sweep from the unit-filling MI state.
As shown in Fig.~\ref{fig:4}(b), in the unfrustrated regime ($J_{\rm fin} > 0$), a macroscopic condensate clearly forms at the $\Gamma$ point.
Figure~\ref{fig:4}(d) shows that the condensate fraction, defined as $S_{\Gamma}/N$, increases with $J_{\rm fin}/U$.
In contrast to the kagome lattice, this macroscopic condensation remains robust even for large system sizes, such as $N = N_{\rm s}=L^2=36^2$.
Furthermore, in Fig.~\ref{fig:4}(c), the semiclassical method successfully captures finite BEC peaks at ${\bf k} = (4\pi/3d_{\rm lat})(1/2,\sqrt{3}/2) \equiv {\bf Q}_{\rm K} $ and ${\bf k} = (4\pi/3d_{\rm lat})(1,0) \equiv {\bf Q}_{\rm K'}$, corresponding to chiral SF order as expected for $J_{\rm fin} < 0$.
This result is consistent with both cluster mean-field calculations~\cite{yamamoto2020frustrated} and recent experiments with $^{39}\text{K}$ atoms~\cite{braund2024negative}.
As shown in Fig.~\ref{fig:4}(e), the condensate fraction at the $K$ and $K'$ points is significantly suppressed with increasing system size.

The corresponding real-space correlation functions, defined as ${\overline C}(r) = {\cal N}(r)^{-1}\sum_{|| {\bf R} - {\bf R}' ||_2 = r} C_{{\bf R}, {\bf R}'}$, are shown in Figs.~\ref{fig:4}(f) and \ref{fig:4}(g).
For $2 \lesssim J_{\rm fin}/|J_{\rm c}| \lesssim 10$, the correlations exhibit a power-law decay, ${\overline C}(r) \approx r^{-1/2}$, 
whereas for $J_{\rm fin}/|J_{\rm c}| \lesssim 2$, they decay exponentially as ${\overline C}(r) \approx e^{ -r / \xi_{\rm cor}}$.
Both regimes can be uniformly described by a modified exponential form: $G(r; \xi_{\rm cor}, \eta) = r^{- \eta} e^{-r/\xi_{\rm cor}}$. 
Our numerical analysis shows that the extracted correlation length $\xi_{\rm cor}$ increases significantly with $J_{\rm fin}/|J_{\rm c}|$, reaching values as large as $\xi_{\rm cor} / d_{\rm lat} \approx O(100)$, while the exponent $\eta$ remains nearly constant around $\eta \approx 0.5$ for $5 \lesssim J_{\rm fin}/|J_{\rm c}| \lesssim 10$~\cite{sm}. 
For $J_{\rm fin} < 0$, the correlation function exhibits a damped oscillatory form, ${\overline C}(r) \approx \cos(\pi \alpha r)e^{-r/\xi_{\rm cor}}$, with an estimated oscillation period $\alpha d_{\rm lat} \approx 1.25$ and correlation length $\xi_{\rm cor} \approx 2 d_{\rm lat}$ in the range $-10 \lesssim J_{\rm fin}/|J_{\rm c}| \lesssim -2$~\cite{sm}.
This oscillatory behavior reflects a short-range 120-degree or chiral order of bosonic phases in the classical limit~\cite{yamamoto2020frustrated}, which is suppressed on average due to the presence of initial quantum fluctuations.

The above results for the triangular lattice serve to highlight key features of the kagome-lattice case. 
First, even in the unfrustrated regime, one-body correlations in the kagome lattice remain remarkably short ranged. 
Second, in the frustrated regime, the lowest band shows no specific condensation, in sharp contrast to the $K$-point condensation observed in the triangular lattice.  
The correlation length is also significantly suppressed compared to that of the triangular lattice. 
The global fraction $F$ over the entire BZ thus provides a useful indicator for distinguishing the short-range correlated state for negative hopping from the MI state, in which $F$ must be equal to $1/3$.

In summary, we have studied the adiabatic sweep dynamics of Bose gases in a kagome optical lattice to elucidate how the sign inversion of hopping amplitudes affects the formation of one-body correlations in the final state.
For positive hopping, an initially prepared Mott insulating state evolves into a $\Gamma$-point superfluid state with a finite condensate fraction. 
In contrast, when the sign of the hopping is inverted, the Mott insulator melts into a disordered state, where the lowest flat band becomes dominantly occupied without condensation into any particular momentum mode, and no long-range order emerges in the one-body correlations. 
These results are qualitatively consistent with recent experiments using ultracold Bose gases in kagome optical lattices~\cite{donini2025bose}. 
Since our semiclassical method captures essential differences between frustrated and unfrustrated regimes, it provides a powerful tool for exploring novel quantum states and nonequilibrium dynamics in geometrically frustrated systems.

\begin{acknowledgments}

We thank Luca Donini and Ulrich Schneider for valuable discussions. 
Numerical simulations were performed on the HOKUSAI supercomputing system at RIKEN and the supercomputer Fugaku provided by the RIKEN Center for Computational Science.
This work was partially supported by the Project No.~JPNP20017, funded by the New Energy and Industrial Technology Development Organization (NEDO), Japan. 
We acknowledge the support from JSPS KAKENHI [Grant Nos.~JP21H04446, JP21H05185, JP23K25830, JP24K06890, JP25K17321]
from the Ministry of Education, Culture, Sports, Science and Technology (MEXT), Japan; 
Quantum Leap Flagship Program from MEXT [Grant No.~JPMXS0118069021]; 
PRESTO from Japan Science and Technology Agency (JST) [Grant No.~JPMJPR245D]; 
FOREST from JST [Grant No.~JPMJFR202T]; 
and ASPIRE from JST [Grant No.~JPMJAP24C2].
We also appreciate funding from 
JST COI-NEXT [Grant No.~JPMJPF2221] and  
the Program for Promoting Research of the Supercomputer Fugaku [Grant No.~MXP1020230411] from MEXT, Japan.  
Further support was provided by 
the RIKEN TRIP initiative (RIKEN Quantum) and
the COE research grant in computational science from Hyogo Prefecture and Kobe City through the Foundation for Computational Science.

\end{acknowledgments}

\bibliography{ref}

\clearpage


\renewcommand{\thetable}{S\arabic{table}} 
\renewcommand{\thefigure}{S\arabic{figure}}
\renewcommand{\thetable}{S\arabic{table}}
\renewcommand\theequation{S\arabic{equation}}
\setcounter{table}{0}
\setcounter{figure}{0}
\setcounter{equation}{0}

\onecolumngrid

\appendix
\begin{center}
\large{Supplemental Material for}\\
\textbf{``
Melting of a Bosonic Mott Insulator in Kagome Optical Lattices with Sign-Inverted Hopping
"}
\end{center}

\section{Details on the discrete truncated Wigner approximation (dTWA) for state-restricted boson models}

The SU($n$) dTWA method provides a systematic framework for incorporating leading-order quantum fluctuations beyond the mean-field dynamics of individual SU($n$) spins~\cite{polkovnikov2010phase,nagao20213}.
Numerical simulations within this framework proceed as follows. 
First, a random configuration of classical spin variables is sampled from a distribution function representing the initial quantum density operator ${\hat \rho}_0$ at time $t=0$.
Second, each sampled configuration is propagated along a classical trajectory governed by a nonlinear equation of motion, derived from the Hamilton's equation at the mean-field level. 
These two steps are repeated many times, similar to Monte Carlo sampling, to generate an ensemble of independent trajectories. 
Within the approximation, the quantum expectation value of an observable is estimated as the ensemble average of the corresponding $c$-number function evaluated over these trajectories.

To apply the SU($n$) dTWA method to the Bose-Hubbard systems, we truncate the local Hilbert space of bosons to a finite dimensional subspace defined as ${\cal H}_{{\rm SU}(n_{\rm max} + 1)} = \text{Span}\{\ket{0},\ket{1},\cdots,\ket{n_{\rm max}}\}$. 
When $n_{\rm max} = 2$ is chosen, the Bose-Hubbard Hamiltonian is effectively mapped to a spin-1 model. 
This truncation is crucial, as the resulting local interactions can be readily linearized using traceless Hermitian matrices from the fundamental representation of the SU(3) group~\cite{davidson20153,nagao20213}.
Applying dTWA in this reduced Hilbert space, the approximated unitary time evolution of the density operator ${\hat \rho}(t) = {\hat U} {\hat \rho}_0 {\hat U}^\dagger$ is expressed as
\begin{align}
{\hat \rho}(t) 
\approx \sum_{ {\bm x}_{1} } \sum_{ {\bm x}_{2} }  \cdots \sum_{{\bm x}_{N_{\rm s}} }  {\cal W}_{0}( {\bm x}_{1},{\bm x}_{2},\cdots,{\bm x}_{N_{\rm s}}  ) \prod_{j}{\hat {\cal A}}^{\rm SU(3)}_{j}[{\bm r}_j(t; {\bm x}_{1},{\bm x}_{2},\cdots,{\bm x}_{N_{\rm s}})],
\end{align}
where $N_{\rm s}$ denotes the total number of sites. 
Here, the initial pseudospin variables ${\bm x}_{j} \in {\mathbb R}^{8}$ at $t=0$ are sampled from the distribution ${\cal W}_{0}( {\bm x}_{1},{\bm x}_{2},\cdots,{\bm x}_{N_{\rm s}})$. 
The local operators ${\hat {\cal A}}^{\rm SU(3)}_{j}[{\bm r}_j(t)] = \frac{1}{3}{\hat 1} + \frac{1}{2}\sum_{\alpha = 1}^{8} r^{\alpha}_{j}(t) {\hat X}^{\alpha}_{j}$ are parameterized by time-dependent eight-dimensional vectors $r^{\alpha}_{j}(t) = r^{\alpha}_{j}(t; {\bm x}_{1},{\bm x}_{2},\cdots,{\bm x}_{N_{\rm s}})$,
where ${\hat X}^{\alpha}_{j}$ are SU(3) generators satisfying the orthogonality condition ${\rm Tr}[ {\hat X}^{\alpha}_{i} {\hat X}^{\beta}_{j} ] = 2 \delta_{i,j} \delta_{\alpha, \beta}$.
The vectors $r^{\alpha}_{j}(t)$, which are nonlinear functions of the initial values ${\bm x}_{j}$, evolve according to a site-decoupled mean-field equation of motion:  
\begin{align}\label{eq:A_SU3_eom}
\frac{d}{dt} {\hat {\cal A}}^{\rm SU(3)}_{j}
&= \frac{i}{\hbar} J(t) [b_{j}, {\hat {\cal A}}^{\rm SU(3)}_{j}] \sum_{j' \in E_{j}(G)}  {\rm Tr}_{j'} \{ {\hat {\cal A}}^{\rm SU(3)}_{j'} b^{\dagger}_{j'} \} + {\rm H.c.} - \frac{i}{\hbar}[{\cal H}^{\rm onsite}_{j}, {\hat {\cal A}}^{\rm SU(3)}_{j}],
\end{align}
where 
$G$ is the graph representing the lattice geometry 
and 
$E_{j}(G)$ denotes the set of neighboring sites connected to site $j$.
The bosonic operators in Eq.~(\ref{eq:A_SU3_eom}) are projected onto the reduced subspace ${\cal H}_{{\rm SU}(3)} = \text{Span}\{\ket{0},\ket{1},\ket{2}\}$ as follows:  
$b_{j} \mapsto \ket{0}\bra{1}_j + \sqrt{2} \ket{1}\bra{2}_j$, 
$b_{j}^{\dagger} \mapsto \ket{1}\bra{0}_j + \sqrt{2} \ket{2}\bra{1}_j$, 
and ${\cal H}^{\rm onsite}_{j} = \frac{U}{2}n_{j}(n_{j}-1) \mapsto U \ket{2}\bra{2}_j$.
Note that Eq.~(\ref{eq:A_SU3_eom}) remains invariant under linear transformations of the generator basis, i.e., $\{ {\hat X}^{\alpha}_{j} \} \rightleftarrows \{ {\hat Y}^{\alpha}_{j} \}$.

If the initial state is a direct product state, the distribution ${\cal W}_{0}$ factorizes into a product of local distributions as ${\cal W}_{0} = \prod_{j=1}^{N_{\rm s}}w_0({\bm x}_j)$.
Each local distribution function $w_0({\bm x}_j)$ is determined as a positive-semidefinite function via a spectral decomposition of the SU(3) generators, following a procedure inspired by quantum state tomography for local quantum states~\cite{zhu2019generalized,nagao20213}. 
Assuming that the initial density operator is a product state, ${\hat \rho}_0 = \prod_{j}{\hat \rho}_0^{(j)}$, the local density matrix at site $j$ can be expressed as 
\begin{align}
{\hat \rho}^{(j)}_0 
&= \sum_{{\bm x}_j} w_{0}({\bm x}_j) {\hat {\cal A}}^{\rm SU(3)}_{j} [ {\bm x}_j ]  = \frac{1}{3}{\hat 1} + \frac{1}{2}\sum_{\alpha = 1}^{8}\overline{x_{j}^{\alpha}} {\hat X}^{\alpha}_{j},
\end{align}
where the overline on $x_{j}^{\alpha}$ denotes the ensemble average over the distribution $w_0({\bm x}_j)$. 
The matrix representation of the generators $\{{\hat X}^{\alpha}_{j}\}$, adopted in Refs.~\cite{davidson20153, nagao20213, nagao2024discrete}, are explicitly represented as  
\begin{align}
{\hat X}_{j}^{1} \mapsto  
\frac{1}{\sqrt{2}}
\begin{bmatrix}
0 & 1 & 0 \\
1 & 0 & 1 \\
0 & 1 & 0
\end{bmatrix}_{j}, \;\;
{\hat X}_{j}^{2} \mapsto  
\frac{1}{\sqrt{2}}
\begin{bmatrix}
0 & -i & 0 \\
i & 0 & -i \\
0 & i & 0
\end{bmatrix}_{j}, \;\;
{\hat X}_{j}^{3} \mapsto 
\begin{bmatrix}
1 & 0 & 0 \\
0 & 0 & 0 \\
0 & 0 & -1
\end{bmatrix}_{j}, \nonumber \\
{\hat X}_{j}^{4} \mapsto 
\begin{bmatrix}
0 & 0 & 1 \\
0 & 0 & 0 \\
1 & 0 & 0
\end{bmatrix}_{j},  \;\;\;\;\;\;\;\; 
{\hat X}_{j}^{5} \mapsto 
\begin{bmatrix}
0 & 0 & -i \\
0 & 0 & 0 \\
i & 0 & 0
\end{bmatrix}_{j}, \;\;\;\;\;\;\;\;\;\;\;\;\;\;\;\;\;\;\;\;\;\;\;\;   \\
{\hat X}_{j}^{6} \mapsto 
\frac{1}{\sqrt{2}}
\begin{bmatrix}
0 & -1 & 0 \\
-1 & 0 & 1 \\
0 & 1 & 0
\end{bmatrix}_{j}, \;\; 
{\hat X}_{j}^{7} \mapsto 
\frac{1}{\sqrt{2}}
\begin{bmatrix}
0 & i & 0 \\
-i & 0 & -i \\
0 & i & 0
\end{bmatrix}_{j}, \;\;
{\hat X}_{j}^{8} \mapsto 
\frac{1}{\sqrt{3}}
\begin{bmatrix}
-1 & 0 & 0 \\
0 & 2 & 0 \\
0 & 0 & -1
\end{bmatrix}_{j}. \nonumber 
\end{align}
Intuitively, the off-diagonal elements of $\{ {\hat X}^{\alpha}_{j} \}_{\alpha \in \{1,2,6,7\}}$ describe the transitions $\ket{0} \rightleftarrows \ket{1}$ and $\ket{1} \rightleftarrows \ket{2}$, 
while those in $\{ {\hat X}^{\alpha}_{j} \}_{\alpha \in \{4,5\}}$ correspond to the creation and annihilation of doubly occupied states.
The diagonal generators ${\hat X}^{3}_{j}$ and ${\hat X}^{8}_{j}$ conserve the onsite occupation number. 
The spectral decomposition of each generator takes the form   
${\hat X}^{\alpha}_{j} = \sum_{s=1}^{3} \lambda^{(j)}_{s} \ket{\phi^{s}_{\alpha}} \bra{\phi^{s}_{\alpha}}_{j}$. 
For the initial Mott insulator (MI) state $\prod_{j}\ket{1}_{j}$, 
the probability of measuring eigenvalue $\lambda^{(j)}_{s}$ of ${\hat X}^{\alpha}_{j}$ is given by $p_{j}(s,\alpha) = | \bra{\phi^{s}_{\alpha}} \ket{1}_{j} |^2 \geq 0$.

\section{Numerical setup for the SU(3) dTWA simulations}

Here, we provide details of the SU(3) dTWA analysis of the linear-sweep dynamics in the kagome Bose-Hubbard model.
The key components of the numerical setup are outlined below. 

\begin{figure}
\begin{center}
\includegraphics[width=160mm]{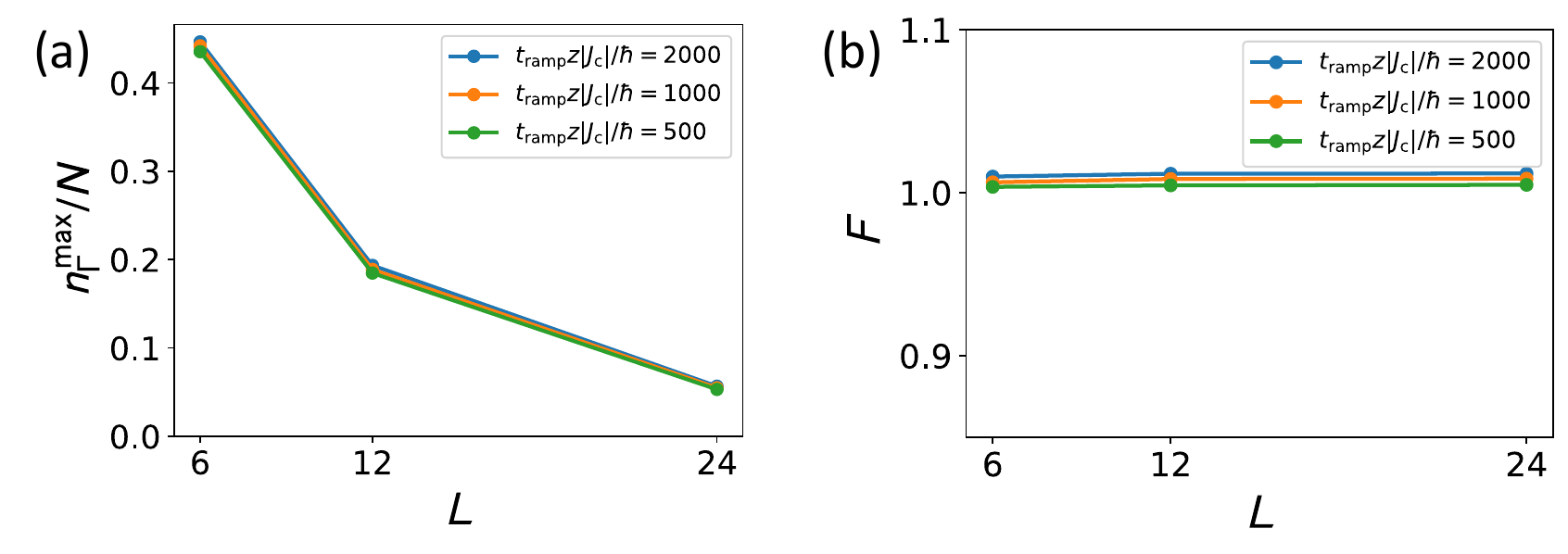}
\vspace{-6mm}
\caption{
Two types of condensate-related fractions in the final state of the kagome Bose-Hubbard model after the sweep, shown for different sweep durations $t_{\rm ramp}$. 
(a)
Condensate fraction at the $\Gamma$ point for the unfrustrated kagome lattice ($J_{\rm fin}>0$) with fixed $g_{\rm fin} = 10$.
The horizontal axis represents the linear system size $L$, with the total number of sites given by $N_{\rm s} = 3 L^2$ and the total particle number $N=N_{\rm s}$.
(b) 
Global flat-band fraction $F = N^{-1} \sum_{{\bf k} \in {\rm B.Z.}} n^{\rm max}_{\bf k}$ over the first Brillouin zone for the frustrated kagome lattice ($J_{\rm fin}<0$), with $g_{\rm fin} = -10$. 
}
\label{fig: supp: convergence}
\end{center}
\end{figure}

\begin{itemize}
\item{
{\bf Differential equation solver}:
Eq.~(\ref{eq:A_SU3_eom}) is implemented as a set of coupled ordinary differential equations for onsite $3 \times 3$ matrix variables defined at each site on the kagome lattice. Simulations are performed under periodic boundary conditions along the ${\bf a}_1$ and ${\bf a}_2$ directions. 
Time evolution is computed using the explicit fourth-order Runge-Kutta method.
}
\item{
{\bf Estimated critical points}:
Within the single-site Gutzwiller approximation, the triangular-lattice Bose-Hubbard model with positive hopping undergoes a quantum phase transition from a MI to a U(1) symmetry-broken superfluid (SF) at zero temperature~\cite{yamamoto2020frustrated}, with the critical value estimated as $U/|J_{\rm c}|^{\rm tri} \approx 35.0$.
When the sign of the hopping is reversed, the ordered phase becomes a chiral SF with $\rm {U}(1) \times {\mathbb Z}_2$ symmetry breaking and $K$-point condensation. In this case, the critical interaction strength is approximately halved, $U/|J_{\rm c}|^{\rm tri} \approx 17.5$.
For the kagome lattice, the critical values are related to those of the triangular lattice via $U/|J_{\rm c}|^{\rm kagome} = (2/3) U/|J_{\rm c}|^{\rm tri}$, yielding $U/|J_{\rm c}|^{\rm kagome} \approx 23.3$ for the unfrustrated case and $\approx 11.7$ for the frustrated case.
}
\item{
{\bf Physical time units}:
To make Eq.~(\ref{eq:A_SU3_eom}) dimensionless, the time $t$ is rescaled by the energy unit $zJ_{\rm c}$, where $z$ is the coordination number of the lattice. 
This choice is natural, as the dTWA equations are derived alongside the site-decoupled Gutzwiller approximation, where the many-body density matrix is approximated by a product of local states. 
}
\item{
{\bf Sweep duration}:
Throughout the main text, 
we set $t_{\rm ramp} z|J_{\rm c}| / \hbar = 2000$, which ensures quasi-adiabatic dynamics with negligible non-adiabatic excitations. 
As shown in Fig.~\ref{fig: supp: convergence}(a), even for shorter sweep durations, the final-state condensate fraction at the $\Gamma$ point converges in the unfrustrated regime. A similar convergence behavior is observed for the global flat-band fraction $F$ in the frustrated regime [Fig.~\ref{fig: supp: convergence}(b)].
}
\item{
{\bf Parallelization and computational resources}:
Trajectory sampling in the dTWA simulations is efficiently parallelized using the Message Passing Interface (MPI) framework~\cite{nagao2024discrete}.
A typical job on the supercomputer Fugaku uses 300 nodes, each launching 40 MPI processes, resulting in a total of $12,000$ concurrent processes. 
Each process evolves a single trajectory initialized with a randomly sampled classical configuration. After time evolution, a representative process collects observables from all trajectories and computes their ensemble averages.
}
\end{itemize}

\section{Momentum distribution of one-body reduced density matrix in the triangular Bose-Hubbard model}

Here, 
we present additional results on the sweep dynamics of the triangular-lattice Bose-Hubbard model. 
The lattice geometry is schematically illustrated in Fig.~\ref{fig:4} of the main text.
The time dependence of the hopping amplitude is the same as in the kagome-lattice case. 
In the triangular lattice, the one-body reduced density matrix (OBRDM) $[ \langle {\hat b}^{\dagger}_{\bf R} {\hat b}_{{\bf R}'} \rangle ]_{{\bf R},{\bf R}'} $ in real space can be diagonalized via Fourier transformation into momentum space.
This transformation yields the atomic momentum distribution $S_{\bf k}$, given by 
\begin{align}
S_{\bf k} = \frac{1}{N_{\rm s}} \sum_{{\bf R},{\bf R}'} \langle {\hat b}^{\dagger}_{\bf R} {\hat b}_{{\bf R}'} \rangle e^{i({\bf R} - {\bf R}')\cdot {\bf k}},
\end{align}
where $N_{\rm s}$ is the total number of lattice sites, and ${\bf R} = m {\hat e}_1 + n  {\hat e}_2$ are the Bravais lattice vectors with lattice spacing $d_{\rm lat}$. 

\begin{figure}
\begin{center}
\includegraphics[width=160mm]{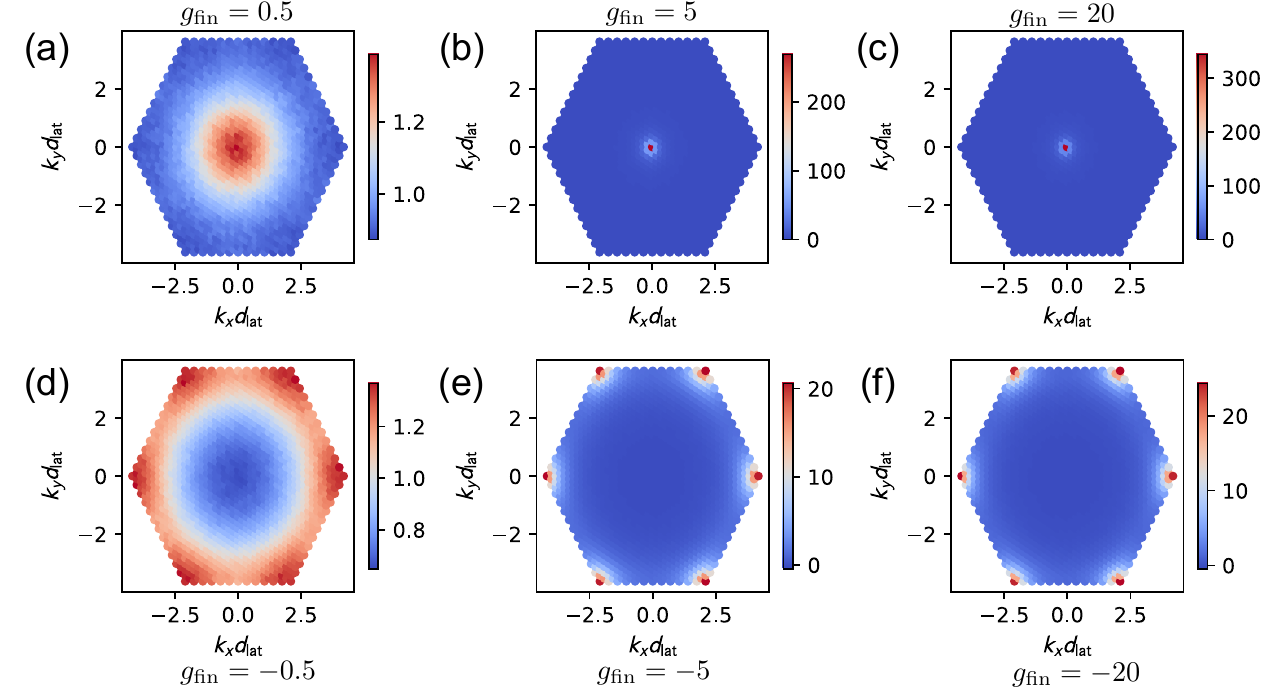}
\vspace{0mm}
\caption{
Dependence of the atomic momentum distribution $S_{\bf k}$ on the final hopping amplitude, evaluated after the sweep in the triangular-lattice Bose-Hubbard model. 
Color bars indicate the magnitude of $S_{\bf k}$.
Panels (a)--(c) show results for the unfrustrated regime with $g_{\rm fin} = 0.5$, $ 5$, and $ 20$, respectively, while panels (d)--(f) show results for the frustrated regime with $g_{\rm fin} = -0.5$, $ -5$, and $ -20$. 
}
\label{fig: supp: tri_mom_dist}
\end{center}
\end{figure}

Figure \ref{fig: supp: tri_mom_dist} shows the atomic momentum distribution $S_{\bf k}$ as a function of the final hopping amplitude $g_{\rm fin}$.
As seen in Figs.~\ref{fig: supp: tri_mom_dist}(a)--\ref{fig: supp: tri_mom_dist}(c), the peak at the $\Gamma$ point grows monotonically with increasing $g_{\rm fin}$. 
The corresponding condensate fraction, defined as $S_{{\bf k} = \Gamma}/N_{\rm s}$, is presented in Fig.~\ref{fig:4} of the main text. 
For negative values of $g_{\rm fin}$, the sign inversion of the hopping amplitude alters the low-energy structure of the system, leading to the emergence of $K$-point condensate at the zone corners as $|g_{\rm fin}|$ increases [Figs.~\ref{fig: supp: tri_mom_dist}(d)--\ref{fig: supp: tri_mom_dist}(f)]. 
The condensation fraction in this regime is defined by summing non-overlapping contributions from the two inequivalent $K$ points: ${\bf Q}_{\rm K} = (4\pi/3d_{\rm lat})(1/2,\sqrt{3}/2) $ and ${\bf Q}_{\rm K'} = (4\pi/3d_{\rm lat})(1,0)$.
Note that for the kagome lattices, the appropriate lattice constant for momentum-space analysis is $2 d_{\rm lat}$, i.e., $d_{\rm lat} \rightarrow 2 d_{\rm lat}$.

\section{Real-space correlation analysis for the triangular-lattice Bose-Hubbard model}

\begin{figure}
\begin{center}
\includegraphics[width=\textwidth,keepaspectratio]{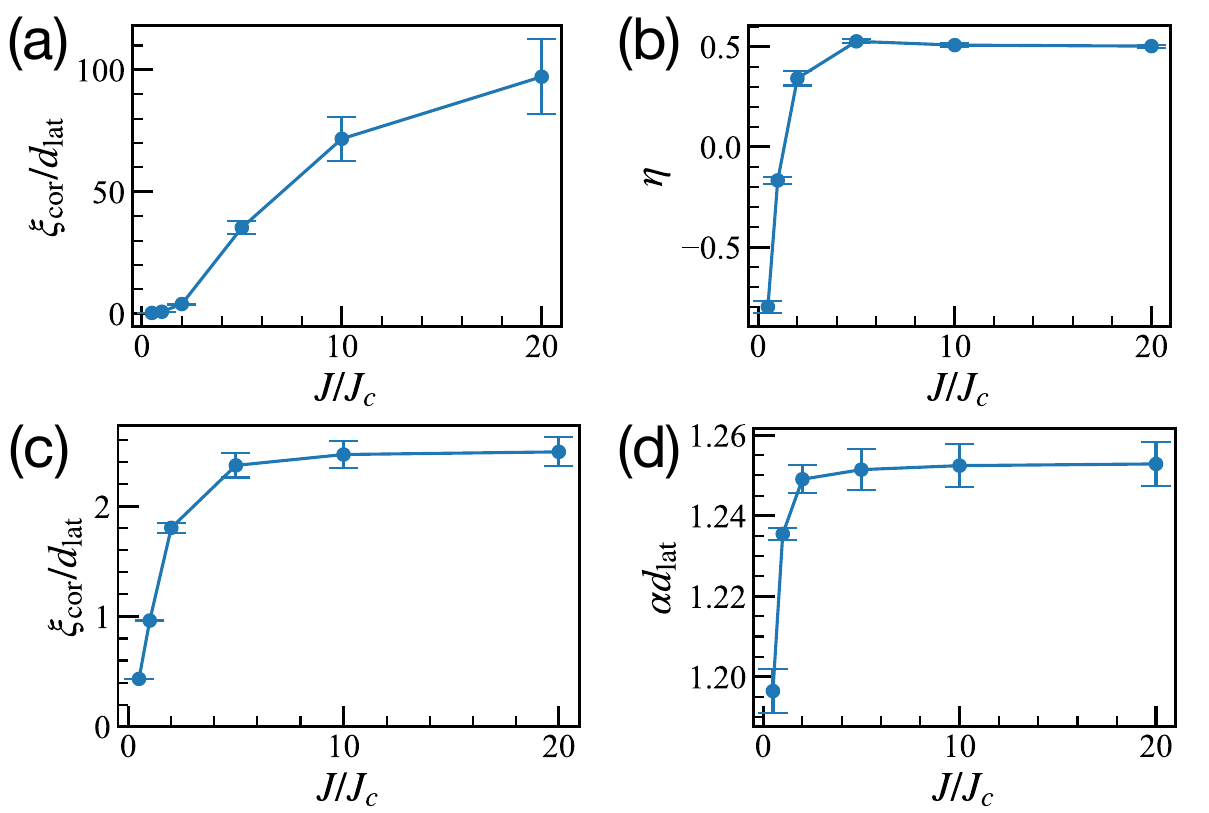}
\vspace{0mm}
\caption{
Fitting results for the real-space correlation functions ${\overline C}(r)$ in the triangular-lattice Bose-Hubbard model. 
(a) 
Correlation length $\xi_{\rm cor}$ extracted using the ansatz $e^{-r/\xi_{\rm cor}} r^{- \eta} d_{\rm lat}^\eta$ in the positive-hopping regime. 
(b) 
Power-law exponent $\eta$ corresponding to panel (a). 
(c)
Same as (a), but for the negative-hopping regime, fitted using the ansatz $\cos(\pi \alpha r) e^{-r/\xi_{\rm cor}}$.
(d)
Oscillation period $\alpha d_{\rm lat}$ corresponding to panel (c). 
All data are obtained for a system size of $N_{\rm s}=36 \times 36$ under periodic boundary conditions.
}
\label{fig: supp_fitdata}
\end{center}
\end{figure}

Here, we supplement the fitting analyses of real-space correlation functions ${\overline C}(r)$ with additional data for the triangular-lattice Bose-Hubbard model. 
In Fig.~\ref{fig: supp_fitdata}(a), we plot the correlation length $\xi_{\rm cor}$, evaluated after the sweep, as a function of $J/J_{\rm c}$ in the positive-hopping regime. We find that $\xi_{\rm cor}$ increases rapidly and can reach values as large as $100 d_{\rm lat}$. 
In contrast, in the negative-hopping regime, the growth of offsite correlations is significantly suppressed due to the combined effects of initial quantum fluctuations and geometrical frustration, as shown in Fig.~\ref{fig: supp_fitdata}(c). 
Figures~\ref{fig: supp_fitdata}(b) and \ref{fig: supp_fitdata}(d) display the power-law exponent $\eta$ and the oscillation period $\alpha$, respectively, both of which appear in the fitting functions introduced in the main text. 
As $J/J_{\rm c}$ increases, $\eta$ saturates around $\eta \approx 0.5$, while the oscillation period stabilizes at approximately $\alpha d_{\rm lat} \approx 1.25$.

\section{Effects of triple occupation: SU(4) dTWA study}

To assess the impact of triply occupied states and to evaluate the reliability of the SU(3) dTWA results on the kagome lattice, we introduce the SU(4) dTWA method as a natural generalization of the approach used in the main text.
We consider the four-dimensional truncated Hilbert space ${\cal H}_{\rm SU(4)} = \text{Span}\{\ket{0},\ket{1},\ket{2},\ket{3}\}$ and approximate the time evolution of the density operator ${\hat \rho}(t)$ as 
\begin{align}
{\hat \rho}(t) 
\approx \sum_{ {\bm x}_{1} } \sum_{ {\bm x}_{2} }  \cdots \sum_{{\bm x}_{N_{\rm s}} }  {\cal W}_{0}( {\bm x}_{1},{\bm x}_{2},\cdots,{\bm x}_{N_{\rm s}}  ) \prod_{j}{\hat {\cal A}}^{\rm SU(4)}_{j}[{\bm r}_{j}(t; {\bm x}_{1},{\bm x}_{2},\cdots,{\bm x}_{N_{\rm s}} )],
\end{align}
where ${\bm x}_{j} = (\cdots, x_{s,s'}^{(j)},\cdots)^{\rm T}$ are real-valued classical SU(4) pseudospin variables. 
The local operators ${\hat {\cal A}}^{\rm SU(4)}_{j}[{\bm r}_{j}(t;{\bm x}_{1},{\bm x}_{2},\cdots,{\bm x}_{N_{\rm s}})]$ represent generalized Bloch vectors corresponding to each SU(4) pseudospins ${\bm r}_{j}(t)$, and can be decomposed as 
\begin{align}
{\hat {\cal A}}^{\rm SU(4)}_{j}[{\bm r}_{j}(t;  {\bm x}_{1},{\bm x}_{2},\cdots,{\bm x}_{N_{\rm s}})] = \frac{1}{4}\sum_{s,s' \in \{0,1,2,3\}} r^{(j)}_{s,s'}(t; {\bm x}_{1},{\bm x}_{2},\cdots,{\bm x}_{N_{\rm s}}) ( {\hat \sigma}_{s} \otimes {\hat \sigma}_{s'} )_{j},
\end{align}
where ${\hat \sigma}_{s}$ ($s=1,2,3$) are the standard Pauli operators and ${\hat \sigma}_{0} = {\hat 1}_{\rm SU(2)}$ is the identity operator on the SU(2) space.
Due to unitarity, we have $r^{(j)}_{0,0}(t) = r^{(j)}_{0,0}(0) =  x^{(j)}_{0,0} = 1$ for all $t$, which ensures the normalization condition ${\rm Tr}( {\hat {\cal A}}^{\rm SU(4)}_{j}[{\bm r}_{j}(t; {\bm x}_{1},{\bm x}_{2},\cdots,{\bm x}_{N_{\rm s}})] ) = 1$.
Each Pauli string ${\hat \sigma}_{s} \otimes {\hat \sigma}_{s'}$ acts on the local SU(4) Hilbert space and has a $4\times4$ matrix representation.
To implement the discrete sampling scheme for the unit-filling MI state $\ket{1}\bra{1}$, we apply a spectral decomposition to the Pauli strings~\cite{zhu2019generalized}.
This yields 
\begin{align}
({\hat \sigma}_{s} \otimes {\hat \sigma}_{s'})_{j} \;\;\; \Rightarrow \sum_{g \in \{0,1,2,3\}} \ket{\phi^{(j)}_{s s'}(g)} \lambda_{g} \bra{\phi^{(j)}_{s s'}(g)}, 
\end{align}
where $\lambda_{g} \in {\mathbb R}$ are the eigenvalues, and the probability to measuring eigenvalue $\lambda_g$ is given by $p^{(j)}_{s s'}(g) =  \left |\bra{\phi^{(j)}_{s s'} (g)} \ket{1} \right |^2$.

\begin{figure}
\begin{center}
\includegraphics[width=170mm]{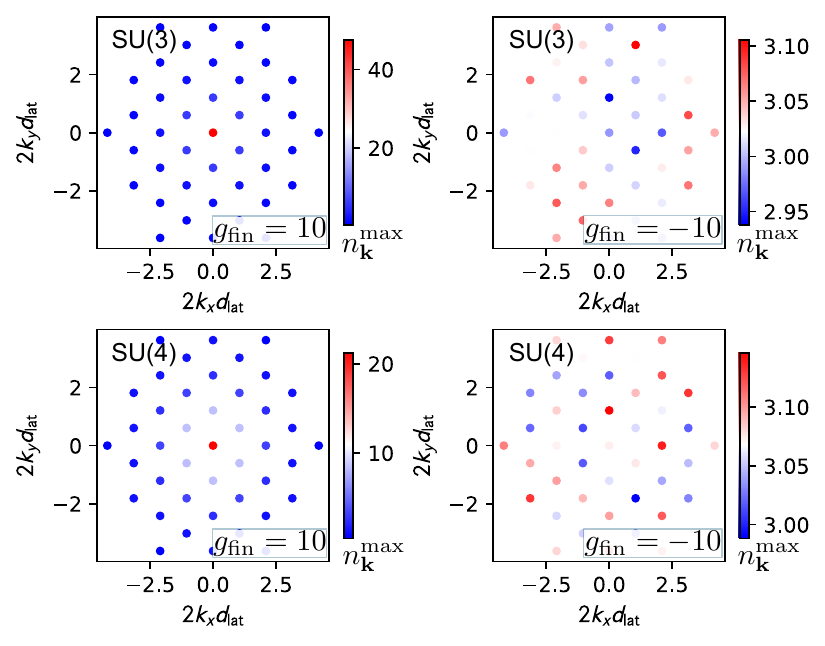}
\vspace{-2mm}
\caption{
Comparison of the momentum distribution of the maximum eigenvalues $n^{\rm max}_{\bf k}$ of the OBRDM, obtained using the SU(3) dTWA (top) and SU(4) dTWA (bottom) for the kagome-lattice Bose-Hubbard model.
The left (right) panels correspond to the unfrustrated (frustrated) regime with $g_{\rm fin} = 10$ ($-10$).
The system size is $N_{\rm s} = 3 \times 6^2 = 108$ with periodic boundary conditions.
All results are evaluated at the final time $t = t_{\rm ramp}$, with the sweep duration set to $t_{\rm ramp} = 1000 \hbar/(z|J_{\rm c}|)$. 
}
\label{fig: supp: su_4_eig}
\end{center}
\end{figure}

\begin{figure}
\begin{center}
\includegraphics[width=160mm]{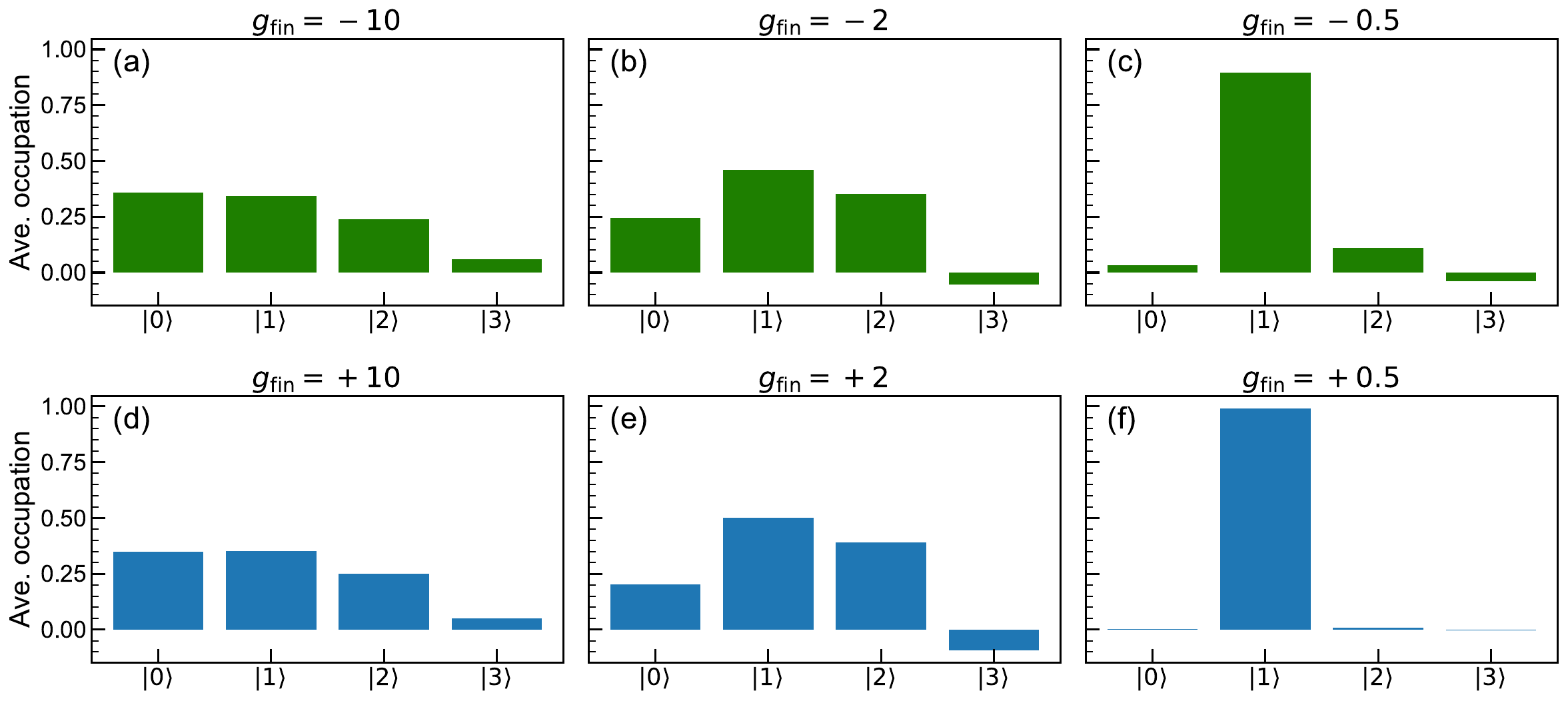}
\vspace{0mm}
\caption{
Occupation probabilities after the sweep with $t_{\rm ramp} = 1000 \hbar/(z|J_{\rm c}|)$, obtained using the SU(4) dTWA for the kagome-lattice Bose-Hubbard model. 
The system size is $N_{\rm s} = 3 \times 6^2$ with periodic boundary conditions.
The vertical axis in each panel indicates the expectation value of the spatially averaged projection operator $N^{-1}_{\rm s} \sum_{j} \ket{n}\bra{n}_{j}$.
Panels (a)--(c) show the results for $g_{\rm fin} = -10$, $-2$, and $-0.5$, respectively, while panels (d)--(f) correspond to $g_{\rm fin} = 10$, $2$, and $0.5$. 
}
\label{fig: supp: histogram_su4}
\end{center}
\end{figure}

\begin{figure}
\begin{center}
\includegraphics[width=160mm]{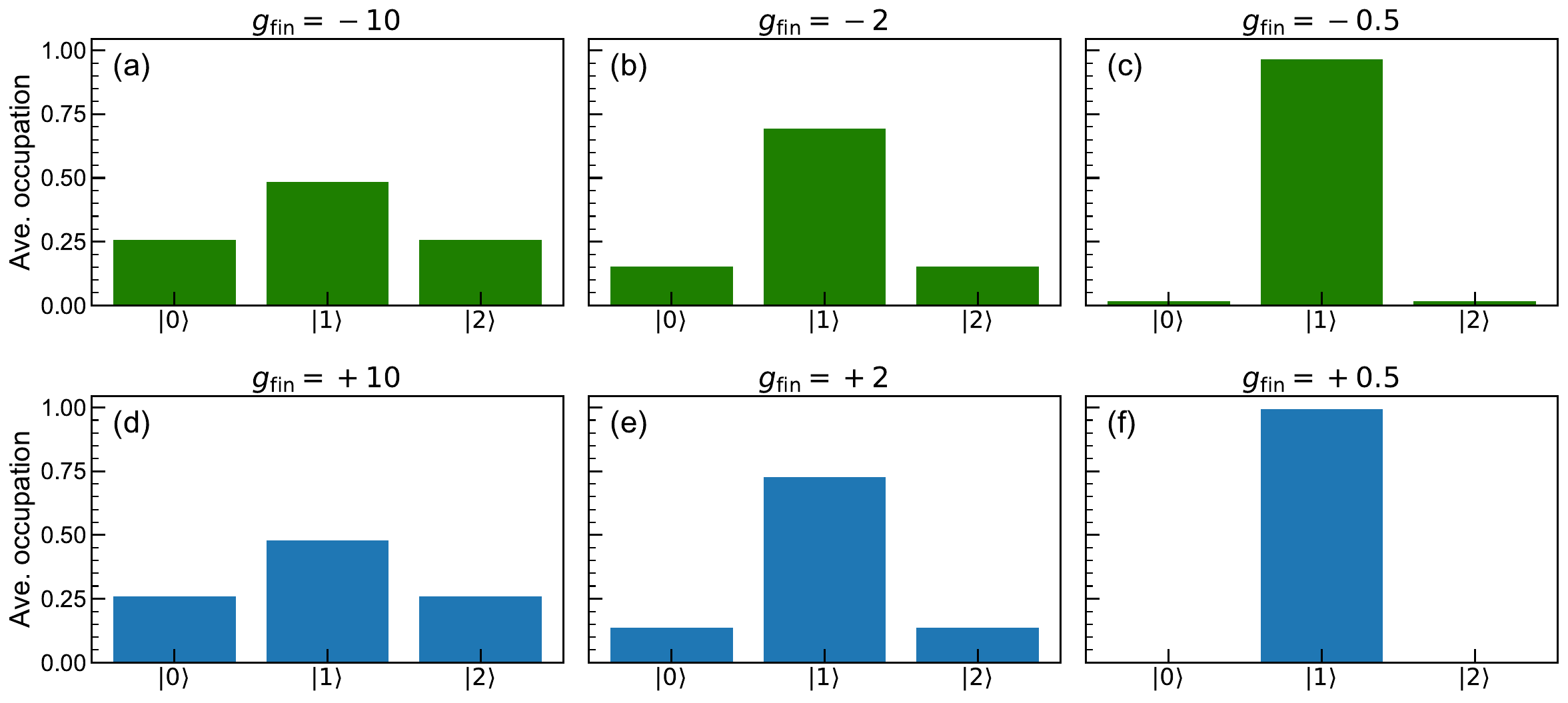}
\vspace{0mm}
\caption{
Same as Fig.~\ref{fig: supp: histogram_su4}, but showing the results obtained using the SU(3) dTWA.
}
\label{fig: supp: histogram_su3}
\end{center}
\end{figure}

Figure~\ref{fig: supp: su_4_eig} presents the momentum distribution of the maximum eigenvalues of the OBRDM, comparing the results obtained with the SU(3) and SU(4) dTWA. 
Specifically, we consider $g_{\rm fin} = \pm 10$, which places the system deeply in the hopping-dominant regime. 
The figure demonstrates that the inclusion of triple occupation does not qualitatively change the behavior in either regime. 
Notably, while the peak height for positive hopping is slightly reduced in the SU(4) case, a clear signature of Bose-Einstein condensation remains visible. 
This reduction can be attributed to changes in the occupation probabilities of the local Fock states, as further analyzed in Figs.~\ref{fig: supp: histogram_su4} and~\ref{fig: supp: histogram_su3}, where we examine the occupation probabilities ${\rm Tr}(\ket{n}\bra{n} {\hat \rho}(t) )$ for each relevant Fock state in the adiabatically prepared states using the SU(4) and SU(3) dTWA, respectively. 
Figures~\ref{fig: supp: histogram_su4}(c) and \ref{fig: supp: histogram_su4}(f) show that when the onsite interaction dominates during the entire sweep (e.g., for $g_{\rm fin} = \pm 0.5$), particle-hole excitations around $\ket{1}$ are strongly suppressed. 
As $|g_{\rm fin}|$ increases, however, the occupation spreads more broadly among the local Fock states. 
For instance, in Figs.~\ref{fig: supp: histogram_su4}(a) and \ref{fig: supp: histogram_su4}(d), with $g_{\rm fin} = \pm 10$ in the SU(4) dTWA, the vacuum state $\ket{0}$ becomes the most populated, followed by $\ket{1}$ and then $\ket{2}$.
In contrast, the SU(3) dTWA results [Figs.~\ref{fig: supp: histogram_su3}(a) and \ref{fig: supp: histogram_su3}(d)] show a dominant occupation at $\ket{1}$, with a nearly symmetric distribution between $\ket{0}$ and $\ket{2}$. This difference accounts for the observed qualitative variations in the momentum distributions. 
Thus, although the population of the triply occupied state $\ket{3}$ remains small in the final state, its presence modifies the population distribution among other local states, as indicated by Fig.~\ref{fig: supp: histogram_su4}.

Finally, we note that the SU(4) dTWA occasionally yields unphysical negative occupation probabilities for the state $\ket{3}$, as seen in Figs.~\ref{fig: supp: histogram_su4}(b) and \ref{fig: supp: histogram_su4}(e).
This issue highlights a persistent limitation in precisely controlling the approximation, despite the SU(4) method's utility in qualitatively capturing triplon-related effects within the semiclassical framework. 
In contrast, the SU(3) dTWA consistently produces physically valid, positive occupation probabilities under our setup, making it more reliable for quantitative analyses.

\end{document}